\documentclass[aip]{revtex4}
\usepackage{setspace, fancyhdr, siunitx, graphicx}
\usepackage{amssymb, nccmath, physics, inputenc}
\usepackage{xparse, subcaption, hyperref}
\newcommand{\lagr}{\mathcal{L}}
\newcommand{\cphi}{\varphi}
\newcommand{\kap}{\kappa}
\newcommand{\br}{\va{r}}
\newcommand{\bv}{\va{v}}

\begin{document}

\title{Galaxy Cluster Statistics in Modified Gravity Cosmologies}

\author{Marcell Howard}
\author{Arthur Kosowsky}
\affiliation{Department of Physics and Astronomy, University of Pittsburgh, Pittsburgh, PA 15260 USA}

\author{Georgios Valogiannis}
\affiliation{Department of Physics, Harvard University, Cambridge, MA 02138, USA}

\begin{abstract}
    If the accelerated expansion of the universe is due to a modification of general relativity at late times, it is likely that the growth of structure on large scales would also display deviations from the standard cosmology. We investigate the statistics of the distribution of galaxy cluster-sized halos as a probe of gravity. We analyze the output of several matched N-body simulations with the same initial conditions and expansion histories but using both DGP and $f(R)$ gravity with various parameters. From each simulation we extract the cluster mass function, power spectrum, and mean pairwise velocity at redshifts $1$, $0.3$, and $0$. All three statistics display systematic differences between gravity theories. The mean pairwise velocity provides an important consistency test for any posited departure from general relativity suggested by measurements of the power spectrum and cluster mass function. Upcoming microwave background experiments, including Simons Observatory and CMB-S4, will detect tens of thousands of galaxy clusters via the thermal Sunyaev-Zeldovich effect, probe their masses with lensing of the microwave background, and potentially measure velocities using the transverse lensing effect or the kinematic Sunyaev-Zeldovich effect. These cluster measurements promise to be a substantial probe of modified gravity.
\end{abstract}

\maketitle

\section{Introduction}

The most common explanation for the current accelerating expansion of the universe is some undetected stress-energy component with negative pressure and dominant energy density. However, no compelling ideas for this component have emerged. A simple dimensional analysis argument for the natural size of any vacuum energy component combines the fundamental constants $c$, $\hbar$, and $G$ to obtain an energy density which is too large by the infamous 120 orders of magnitude. Dynamical components such as a scalar field face similar problems. Within the framework of general relativity, accelerating cosmic expansion today requires new physics ("Dark Energy") which dominates the energy budget of the universe with a characteristic energy scale corresponding roughly to room temperatures -- a scale which is constrained by innumerable experimental results.

This conceptual difficulty has led to consideration of the only other logical alternative: that general relativity (GR) is not the correct theory to describe cosmological dynamics (for a comprehensive review, see \cite{Clifton12}). general relativity has passed numerous stringent observational tests, but all of these are on length scales corresponding to the size of our solar system or smaller. It is therefore a possibility that a fundamental theory of gravity mimics general relativity on small scales, while having a different form on scales much larger than the solar system. Constructing such theories is challenging, because any modification must be made small enough to be consistent with current measured values of the parameterized post-Newtonian (PPN) parameters \cite{Will14}. The propagation speed of gravitational waves being that of light to within one part in $10^{16}$ \cite{Abbott16, Abbott17} also rules out many possible alternate gravity theories \cite{Langlois18}. 

Gravity has two cosmological effects: it determines the expansion history of the universe given its energy-momentum content, and it drives the growth of primordial perturbations into cosmological structures. The combination of these effects, along with initial conditions visible in the anisotropies of the cosmic microwave background, determines the detailed distribution of density perturbations at late times on large scales. Any modification of GR on large scales therefore will, to some extent, affect the distribution of matter on large scales. 

Here we investigate the impact of differing gravitational theories on the largest bound masses in the universe, corresponding to galaxy clusters. Galaxy clusters are rare objects which form in locations corresponding to rare peaks in the initial mass distribution. The mass of a given individual halo is determined largely by gravitational growth moderated by the cosmic expansion. At a finer level of detail, tidal interactions and mergers with other nearby halos also contribute to the present halo mass. (Radiation feedback from the formation of stars and active galactic nuclei can also have an effect, but the largest galaxy clusters have a deep enough gravitational potential that the effect appears to be minor \cite{Bocquet15}; we ignore all baryonic effects in this work.) Both the growth factor and the expansion history can be altered in modified gravity compared to general relativity, as can merger histories and tidal effects. 

While it is not possible to point to properties of an individual cluster as reflecting modified gravity, the statistical distribution of observable cluster properties reflects the underlying theory of gravity. In this work, we analyze a number of matched cosmological N-body simulations with identical initial conditions and differing gravitational physics to quantify the statistical differences in galaxy cluster populations due to the nature of gravity. We first demonstrate the close correspondence between the largest-mass halos in all simulations, which arises due to the identical initial conditions. We then extract the distribution of large halo masses, the halo spatial power spectrum, and the mean pairwise halo velocity for each simulation and compare with those from the standard cosmology with general relativity. Systematic differences ranging from a few percent to tens of percent are evident in the simulations considered. 

Observationally, galaxy cluster statistics are a promising route to constraints on gravity. In recent years, cosmic microwave background experiments including the Planck Satellite \cite{Planck15}, the Atacama Cosmology Telescope \cite{SO19}, and the South Pole Telescope \cite{Bleem20} have mapped large sky regions at angular resolution and sensitivity sufficient to detect thousands of galaxy clusters via the thermal Sunyaev-Zeldovich (tSZ) effect. The tSZ effect, which creates a spectral distortion in the microwave radiation passing through the galaxy cluster's ionized gas, is essentially independent of cluster redshift and gives a relatively simple selection function for clusters \cite{Bleem20, Hilton21}. Once identified, cluster positions can be determined in three dimensions through follow-up galaxy redshift observations; cluster masses can be measured using optical weak gravitational lensing, such as \cite{Murata19}, or statistically using lensing of the microwave background \cite{Madhavacheril15}; and cluster velocities can potentially be obtained from the kinematic Sunyaev-Zeldovich effect \cite{Calafut21, Schaan21} or potentially from the transverse lensing effect \cite{Hotinli19}. As we anticipate a new generation of microwave experiments with steadily increasing sensitivity, quantifying the possibilities for constraining modified gravity is of particular interest. These results can also be used to calibrate approximate calculations of structure growth in modified gravity theories such as the mass peak-patch formalism \cite{Stein18, Stein20}.

The paper is organized as follows: in section 2, we briefly review the Hu-Sawacki \cite{Hu07} and (normal-)DGP \cite{Dvali00} modified gravity models and their effects on cosmological scales. Section 3 summarizes the cosmological simulations on which our analysis is based, and establishes the correspondence of massive halos between the different simulations. Sections 4, 5, and 6 present the cluster halo mass function, cluster power spectrum, and halo mean pairwise velocity, and compare between the different models. A summary and outlook section discusses extending these results to more general parameterizations of modified gravity, the potential of upcoming data to measure these cluster statistics well enough to constrain modified gravity, and the relation of cluster statistics to other probes of large-scale structure. 

\section{Modified Gravity Models}

\subsection{Hu-Swacki $f(R)$ Gravity}

This work focuses on scalar-tensor theories of gravity, with extra spin-zero degrees of freedom added to the Einstein-Hilbert gravitational action. In particular, we consider two definite models. The first is $f(R)$-gravity, where the Ricci curvature scalar in the action is replaced with a general function of the curvature scalar. In particular, the Hu-Sawicki model \cite{Hu07} is given by (in units of $c = 1$)
\begin{equation}
    S = \frac{1}{2\kap^2}\int\dd[4]{x}\sqrt{-g}(R + f(R)) + S_m,
\end{equation}
where $S_m$ is the action for any additional matter fields, the reduced four-dimensional Planck mass is $\kap^{-1} \equiv \frac{1}{\sqrt{8\pi G}} \approx \SI{2.43e18}{\giga\electronvolt}$, and
\begin{equation}
    f(R) \equiv -\mu^2 \frac{c_1(R/\mu^2)^n}{1 + c_2(R/\mu^2)^n},
\end{equation}
where $\mu^2 \equiv \Omega_m H_0^2$ with $\Omega_m$ being the present time fractional energy density for matter and $H_0$ is the value of the Hubble parameter today. In its original inception, the model was defined by three parameters $c_1$, $c_2$, and $n$. At large curvature such that $\abs{\mu^2/R} \ll 1$, we get
\begin{equation}
    f(R) \approx -\frac{c_1}{c_2}\mu^2 + \frac{c_1}{c_2^2}\mu^2\qty(\frac{\mu^2}{R})^n.
\end{equation}
We take $n = 1$, a simple special case which captures the qualitative behavior of the model. To reproduce accelerated expansion, the ratio $c_1/c_2$ is set to recover the effective cosmological constant in the high-curvature limit: 
\begin{equation}
    f(R) \approx -6\Omega_\Lambda H_0^2 + |f_{R0}|\frac{\bar{R}^2}{R},
\end{equation}
where $\bar{R}$ is the background curvature value evaluated in $\Lambda$CDM today
\begin{equation}
    \bar{R} = 3\Omega_{m}H_0^2\qty(1 + 4\frac{\Omega_{\Lambda}}{\Omega_{m}}),
\end{equation}
$f_{R0}$ is given by \cite{Valogiannis19}
\begin{equation}
    {f}_{R0} \equiv \eval{\dv{f}{R}}_{R={\bar R}} = -\frac{c_1}{c_2^2}\qty(\frac{\Omega_{m}}{3(\Omega_{m} + 4\Omega_{\Lambda})})^{2},
\end{equation}
and $f_R \equiv \dv{f}{R}$ functions as a new scalar degree of freedom termed the "scalaron". Notice in the limit of $c_1/c_2^2\rightarrow0$, we recover the standard $\Lambda$CDM cosmology, meaning we can parameterize deviations from GR solely by varying the magnitude of $f_{R0}$. 

We consider three specific models by choosing $\abs{{f}_{R0}} = \{10^{-6}, 10^{-5}, 10^{-4}\}$, which will be referred to as F6, F5, and F4 respectively (which we have ordered by increasing deviation from GR). The Hu-Sawicki theory is an example of a class of modified gravity theories that possesses a chameleon mechanism or a mechanism which screens the effects of modified gravity in regions where the density exceeds a characteristic threshold. This is equivalent to imparting a spatial dependence on the mass of the scalaron i.e.
\begin{equation}
    m^2(x) = \frac{1}{3}\qty(\frac{1 + f_R}{f_{RR}} - R) \approx \frac{1}{3f_{RR}} \equiv \frac{1}{3}\abs{\dv[2]{f}{R}}^{-1},
\end{equation}
where we took the high curvature limit. As a result, a fifth force with a Yukawa-like potential is introduced which acts on all massive particles \cite{Alam21}. The modified Poisson's Equation (in the quasi-static limit) is
\begin{equation}
    2\nabla^2\Phi = \kap^2a^2\var{\rho_m} + \nabla^2f_R,
\end{equation}
where $\Phi$ is the gravitational potential in the conformal Newtonian gauge, $\var{\rho}_m$ is the perturbation in the matter energy density and in the $f(R)\rightarrow 0$ limit, we get the usual unperturbed Poisson's Equation. The perturbed field satisfies the constraint
\begin{equation}
    -3\frac{\nabla^2f_R}{a^2} = \kap^2\var{\rho}_m + \var{R},
\end{equation}
where $\var{R}$ is the perturbed Ricci scalar and the dynamics of this scalar results in the modified Friedman Equation
\begin{equation}
    H^2 + \frac{1}{6}f - (\dot{H} + H^2)f_R + 6H(\ddot{H} + 4H\dot{H})f_{RR} = \frac{\mu^2}{3},
\end{equation}
 where overhead dots denote derivatives with respect to time. The new terms involving $f$ and its derivatives are negligible compared to the Hubble parameter $H^2$, which results in the background expansion to be approximately the same as it is in $\Lambda$CDM. 
 
 \subsection{DGP Gravity}

The second model we consider is the normal DGP (or nDGP as its more commonly known) gravity \cite{Dvali00, Lombriser09} which posits that our four dimensional spacetime is embedded within a five dimensional brane. The action for the theory is 
\begin{equation}
    S = \int\dd[4]{x}\sqrt{-g}\qty(\frac{1}{2\kap^2}R + \lagr_m) + \frac{1}{2\kap^2r_c}\int\dd[5]{x}\sqrt{-g_5}R_5,
\end{equation}
where $r_c$ (called the Vainshtein radius) is the ratio between the 5D and 4D Planck mass \cite{Lombriser09} which acts as a free parameter that determines the length-scale for which we recover standard $\Lambda$CDM cosmology, and $R_5$ and $g_5$ are the five-dimensional curvature scalar and metric determinant respectively. The equations of motion are
\begin{equation}
    2\nabla^2\Phi = \kap^2a^2\var{\rho}_m + \nabla^2\cphi,
\end{equation}
\begin{equation}
    \nabla^2\cphi + \frac{r_c^2}{3a^2\beta(a)}\qty[(\nabla^2\cphi)^2 - (\nabla_i\nabla_j\cphi)^2] = \frac{\kap^2a^2}{3\beta(a)}\var{\rho}_m,
\end{equation}
where $\Phi$ is the gravitational potential in the conformal Newtonian gauge and $\beta(a)$ is
\begin{equation}
    \beta(a) = 1 + 2Hr_c\qty(1 + \frac{\dot{H}}{3H^2}).
\end{equation}
In the limit $\cphi\rightarrow0$ (equivalent to taking the $r_c\rightarrow\infty$ limit in $\beta(a)$), we recover the standard Poisson's Equation for the Newtonian gravitational potential. The modified Friedman equation is given by \cite{Valogiannis19}
\begin{equation}
    H^2 + \frac{1}{r_c}H = \frac{\kap^2}{3}\rho_{\rm tot},
\end{equation}
where $\rho_{tot}$ contains the contributions from radiation, matter, and the cosmological constant. In the $r_c \rightarrow\infty$ limit, we recover the usual Friedman Equation. The extra term has the effect of slowing the expansion and thus we must put in by hand the cosmological constant in order to replicate late-time acceleration \cite{Lombriser09}. We set $H_0r_c = 1$ or $H_0r_c = 5$ and refer to these two models as N1 and N5 respectively (here N1 is the stronger deviations from GR and N5 is the weaker deviation). nDGP is an example of a theory that exhibits the Vainshtein mechanism i.e. the mechanism in which the effects of modified gravity are screened at scales smaller than a characteristic length scale (the Vainshtein radius).

We emphasize here that because modifications to GR are considered small, they are treated as perturbations on top of the background expansion.

\begin{figure}
    \centering
    \includegraphics[width=8.6cm]{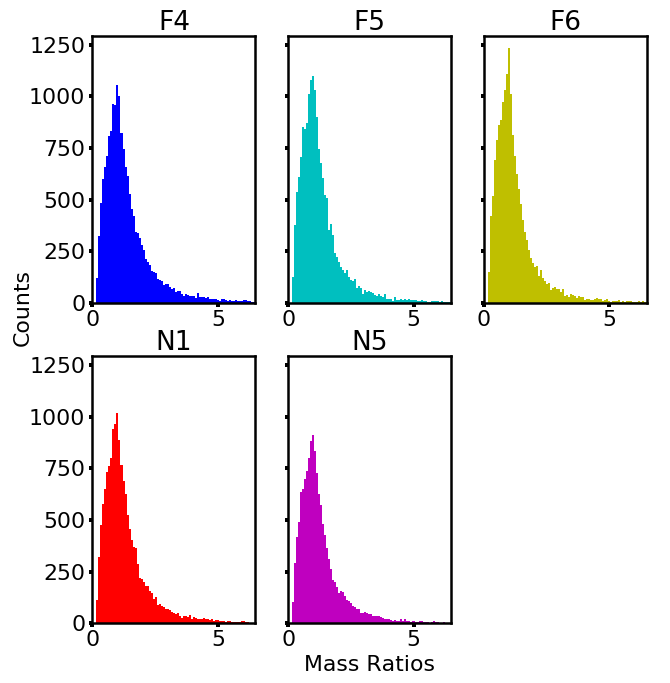}
    \caption{The number of halos in that ratio bin versus The ratio of the mass between a modified gravity halo and its corresponding matched pair in GR (at $z = 1$) under the constraint that the MG halo mass is at least 20\% of the GR halo's mass.}
    \label{MD}
\end{figure}

\begin{figure}
    \centering
    \includegraphics[width=8.6cm]{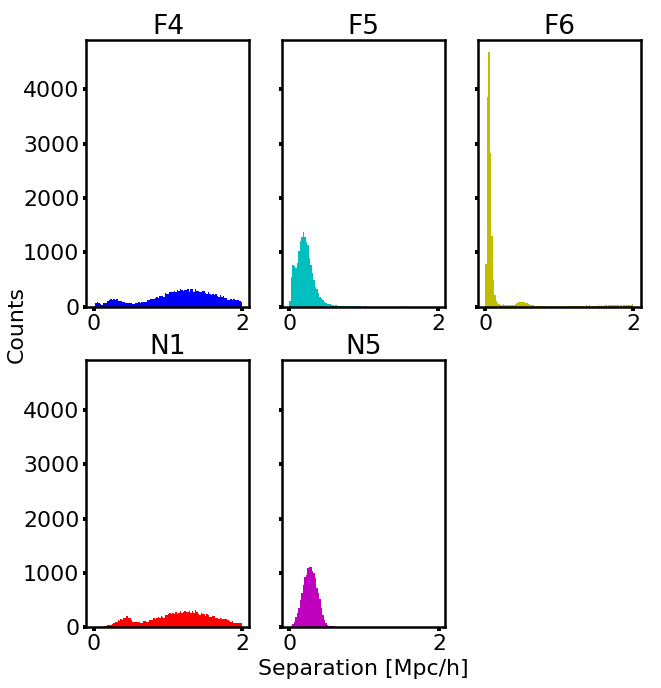}
    \caption{On the horizontal axis is the distance that the modified gravity halo has traveled from its matched halo GR pair. The vertical axis is the number of halos in a given separation bin.}
    \label{SD}
\end{figure}

\begin{figure}
    \centering
    \includegraphics[width=8.6cm]{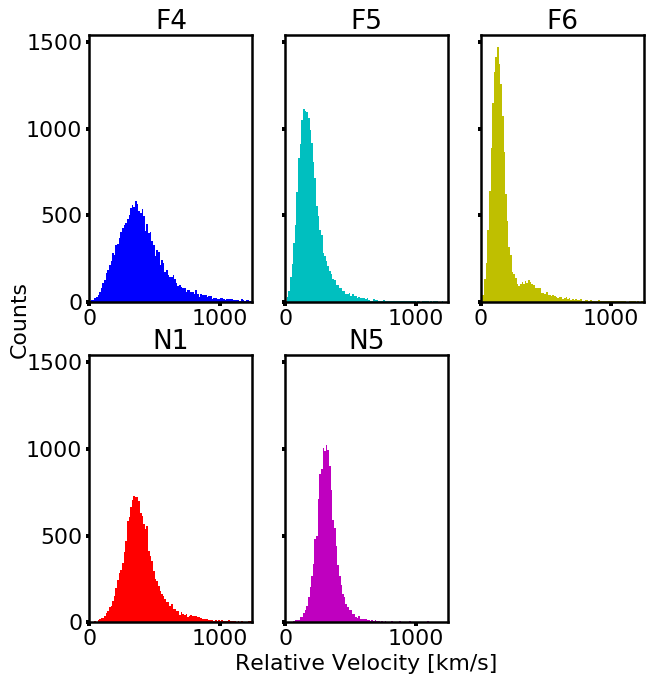}
    \caption{The magnitude of the relative velocity between a modified gravity halo and its matched GR pair for the same set as halos as in Figure \ref{MD}.}
    \label{RSD}
\end{figure}

\section{The \emph{ELEPHANT} Simulations and \emph{ECOSMOG} \& \emph{ECOSMOG-V}}

We extract statistical properties of cluster-mass halos from the set of \emph{ELEPHANT} (Extended LEnsing PHysics using ANaltyic ray Tracing) simulations of cosmic structure growth with modified gravity \cite{Cautun18}. These cosmological N-body simulations use the \emph{ECOSMOG} (Efficient COde for Simulating MOdified Gravity) \cite{Li12} for $f(R)$ gravity, and the \emph{ECOSMOG-V} \cite{Barreira15} code for nDGP gravity. Both of these codes are modifications of the N-body and hydrodynamic simulation code \emph{RAMSES} \cite{Teyssier02}. The \emph{ELEPHANT} simulations have been used previously for \cite{Alam21}. The \emph{ECOSMOG} algorithm is adapted specifically to solve the equations of motion for an $f(R)$ theory, particularly theories that incorporate a chameleon mechanism whereas \emph{ECOSMOG-V} is particularly well suited for MG theories that have a Vainshtein mechanism (the V in \emph{ECOSMO-V} stands for Vainstein).

These codes differ from the standard cosmological N-body code in two ways: (1) they evolve density perturbations using a modified Poisson equation reflecting the modified gravity effective force law and (2) they incorporate a screening mechanism to mitigate the effect of modified gravity on small scales. 

Each simulation is contained in a box with side length $L_{\rm Box} = 1024$ Mpc/h and periodic boundary conditions, with $N = 1024^3$ particles of mass $7.78 \times 10^{10} h^{-1} M_\odot$. Each simulation began with the same initial conditions, given by a specific realization of a Gaussian random density field with a power spectrum consistent with the cosmic microwave background anisotropies at early times. Density perturbations are assumed to be adiabatic with equal fractional density perturbations in all components at every point in space. The density power spectrum at the starting redshift $z_i = 49$ can be obtained from standard codes such as \emph{CLASS} \cite{Blas11}, assuming evolution of the initial density field at recombination using the usual linear growth factor from general relativity. In other words, the simulations assume that any modified gravity effects are only relevant at redshifts below $z = z_i$, and general relativity is a good approximation at earlier times. We make use of code outputs at redshifts $z \in \{1, 0.3, 0\}$. We also made use of only one realization (Box 1) out of the five that were available.

Halo catalogs were compiled from the simulations using the halo finder \emph{ROCKSTAR} (Robust Overdensity Calculation using K-Space Topologically Adaptive Refinement) \cite{Behroozi12}. \emph{ROCKSTAR} is a 6D phase-space algorithm to locate dark matter halos, their substructure, and non-spherical distortions. Halo masses are defined as
\begin{equation}
    M_{200c} = \frac{4\pi}{3}\rho_{c}R^3_{200c},
\end{equation}
i.e.\ the amount of mass contained within a sphere of radius $R_{200c}$ enclosing 200 times the critical density. The lowest mass objects in these catalogues are $\order{10^{11}} M_\odot/h$.

The cosmological parameters used for these simulations were the best-fit WMAP9 parameters \cite{Hinshaw13}: $\Omega_b = 0.046$, $\Omega_{\rm CDM} = 0.235$, $\Omega_\Lambda = 0.719$, $h = 0.697$, $n_s = 0.971$, and $\sigma_8 = 0.820$. The \emph{ROCKSTAR} catalogue includes masses, co-moving coordinates, and peculiar velocities which we use in this analysis. 

\section{Halo Matching}

In order to characterize the differences between GR and a modified gravity model, we first find corresponding objects in the two simulations. First, we collect all GR halos with a mass of $M > 10^{14} M_\odot/h$ ($N = 18138$). We focus on this mass range because upcoming high-resolution CMB experiments will plausibly detect individual clusters in this mass range via the thermal SZ effect. The most massive MG halo within a co-moving distance of 2 Mpc/h from the GR halo is considered the match of the GR halo, if the MG halo mass is within a factor of 5. In the simulations, every GR halo had at least one such matching in each MG simulation. A small fraction of the GR halos (759 out of 18138, or 4.18\%) had more than one matching halo in at least one of the MG simulations. In these cases, determining the MG match in one or more simulations could prove ambiguous, so for simplicity we simply exclude these GR halos from our analysis. 

All of the matched pairs have a separation less than 2 Mpc/h. This makes sense, because typically a cluster will have a characteristic peculiar velocity of a few hundred km/s (say 300 km/s), and over a cosmological time scale of a few Gyr (say 10 Gyr), it would move by ~3 Mpc. The difference in peculiar velocities between MG and GR will generally be significantly smaller than the GR peculiar velocity, provided that the modification to gravity is small. 

To compare the matched pairs, we looked at three different statistics: the ratio between a halo mass as produced within a MG model to a halo produced within GR, the magnitude of the relative position (which we call the separation) between a MG halo and a GR halo, and the magnitude of the relative velocity between the MG halo and the GR halo. 

Figure \ref{MD} shows the distribution of the ratio between the mass of a MG halo and its matched pair GR halo at a redshift of $z = 1$. While each distribution is peaked near a mass ratio of 1, each MG simulation has some matched halos with substantial mass differences, up to a factor of 5. Since the initial conditions are the same for each model, the discrepant masses reflect physical differences in the evolution of the halos. These likely include differing amounts of subhalos being accreted, and enhanced or depressed tidal disruption. Figure \ref{SD} is the distribution of halo separations between a MG halo and its GR matched pair. MG models that more strongly deviate from GR have a larger spread of pair separations. For models further from GR, the separation distribution also becomes somewhat asymmetric around the mean. Figure \ref{RSD} displays the distribution of the relative velocities between a MG halo and its matched pair GR halo, which appear generally consistent with the spread of displacements.  All of these distributions tend to be slightly skewed towards larger differences, because the modifications of gravity considered all tend to increase the strength of gravity at late times. 

\section{Halo Mass Function}

\begin{figure}
    \centering
	\includegraphics[width=8.6cm]{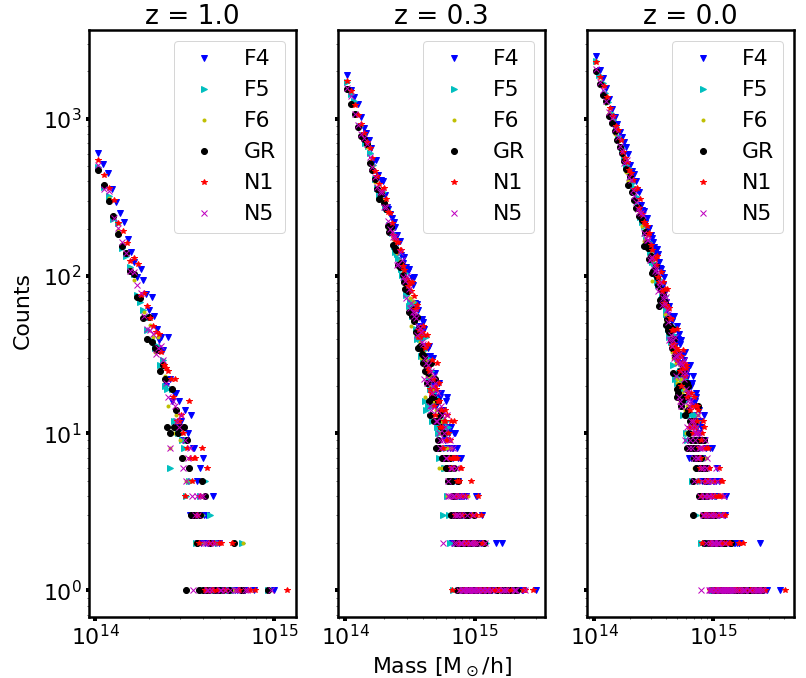}
	\caption{The halo mass function for each gravity model at each of the given redshifts.}
	\label{halo_mass}
\end{figure}

\begin{figure}
    \centering
    \includegraphics[width=8.6cm]{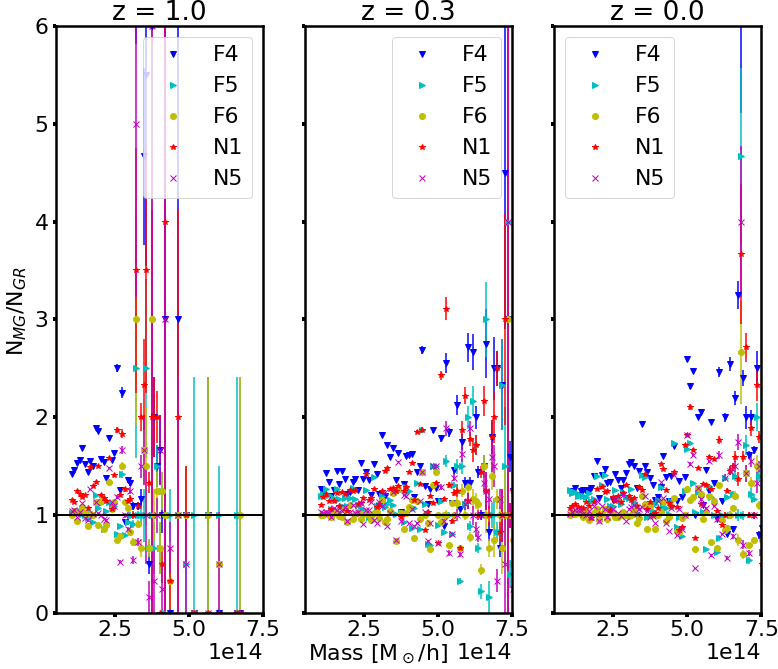}
    \caption{Ratio of the number of halos in each mass bin for a modified gravity model and general relativity.}
    \label{mg_halo}
\end{figure}

The simplest statistic characterizing the cluster halo distribution is the halo mass function, counting the number of halos in various mass bins. Figure \ref{halo_mass} shows the cluster mass function for halos with masses $M_{200c} > 10^{14} M_\odot/h$ at three different redshifts $z=1$, $z=0.3$, and $z=0$, while Fig.~\ref{mg_halo} replots these data showing the ratio of the number of clusters in each modified theory mass bin to the number in the corresponding GR mass bin. Poisson error bars for the number of clusters in each mass bin are shown; these characterize how well the given simulations determine the mass function in each bin. At any redshift, higher-mass bins contain fewer clusters and thus have larger uncertainties in the signal. The number of halos in the largest mass bins increases significantly between $z=1$ and $z=0.3$, but the growth of massive halos is considerably reduced between $z=0.3$ and $z=0$ due to the accelerating expansion of the universe suppressing halo growth compared to the matter-dominated era. 

The models that have the largest increase in gravitational strength from GR (i.e.\ F4 and N1) produce larger numbers of massive halos, as expected. Similar results were shown in Ref.~\cite{Alam21} using the same simulations. This effect appears consistent across redshifts. Halos at the lower mass limit of $10^{14} M_\odot/h$ show significantly less departure from GR than higher-mass halos, as is clear in Fig.~\ref{mg_halo}; this is evidence that the chameleon or Vainshtein screening mechanisms in these models are having significant impact at the lower end of the cluster mass range. 

Cluster halo mass correlates with the easily measured thermal SZ distortion, by which the majority of clusters have been detected. The thermal SZ has a relatively simple selection function which is approximately mass limited. Probing small differences in the mass function will likely require direct mass estimates from either optical weak lensing or CMB lensing. Of the statistics considered here, the mass function has the largest signal but possibly the most complicated systematics to account for. 

\section{Power Spectrum}

\begin{figure}
    \begin{subfigure}[b]{0.66\textwidth}
    \includegraphics[width=8.6cm]{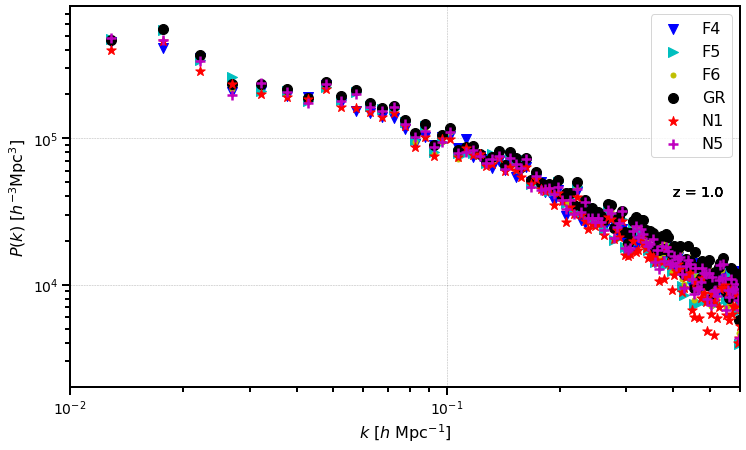}
    \label{fig:ps1}
    \end{subfigure}
    \begin{subfigure}[b]{0.66\textwidth}
    \includegraphics[width=8.6cm]{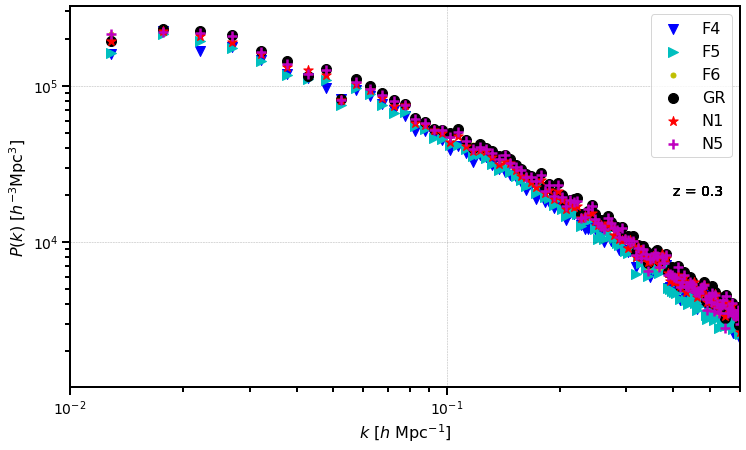}
    \label{fig:ps2}
    \end{subfigure}
    \begin{subfigure}[b]{0.66\textwidth}
    \includegraphics[width=8.6cm]{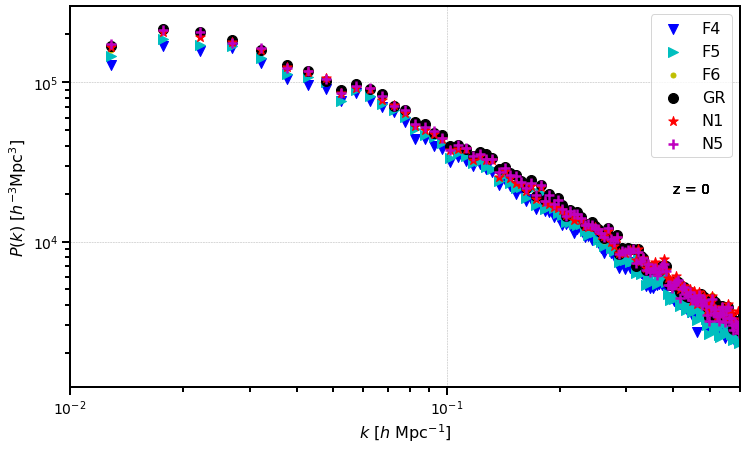}
    \label{fig:ps3}
    \end{subfigure}
    \caption{The power spectrum for each gravity model in order of decreasing redshift plotted over halos with $M_{200c} \geq 10^{14}M_\odot/h$, for redshifts $z = 1$ (top), $z = 0.3$ (middle), $z = 0$ (bottom). All spectra were computed in the publicly available code \emph{nbodykit} \cite{Hand18}.}
    \label{ps_all}
\end{figure}

\begin{figure}
    \begin{subfigure}[b]{0.66\textwidth}
    \includegraphics[width=8.6cm]{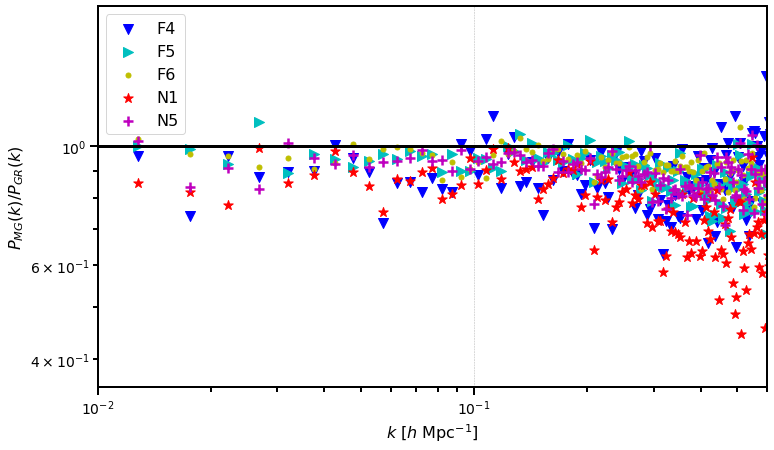}
    \label{psr1}
    \end{subfigure}
    \begin{subfigure}[b]{0.66\textwidth}
    \includegraphics[width=8.6cm]{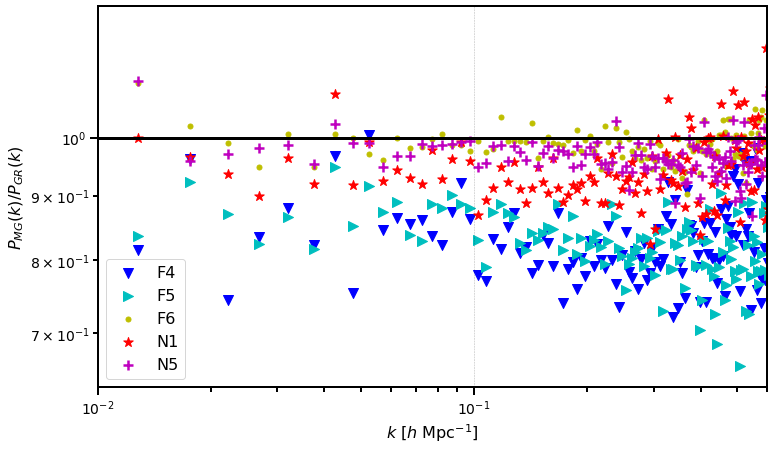}
    \label{psr2}
    \end{subfigure}
    \begin{subfigure}[b]{0.66\textwidth}
    \includegraphics[width=8.6cm]{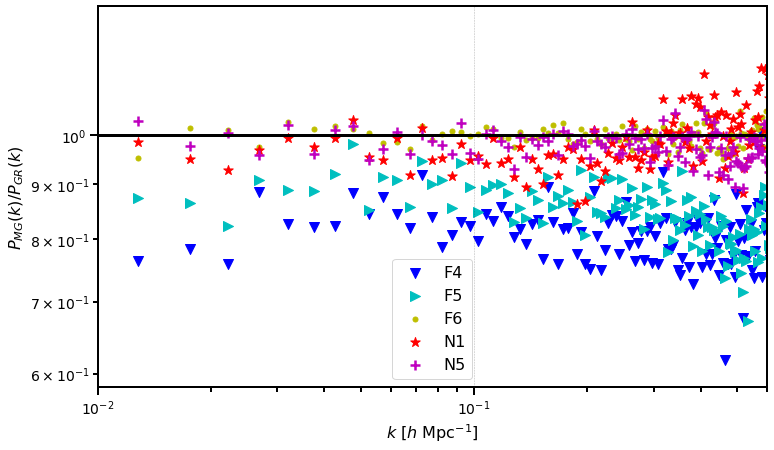}
    \label{psr3}
    \end{subfigure}
    \caption{The ratio of power spectra between a modified gravity model and general relativity for the same mass range as prior, for redshifts $z = 1$ (top), $z = 0.3$ (middle), $z = 0$ (bottom).}
    \label{ps_rat}
\end{figure}

Next we consider the spatial matter power spectrum of the cluster halo distribution. All spectra were computed with the publicly available code \emph{nbodykit} \cite{Hand18}. Figure \ref{ps_all} shows the power spectra for our models at each of the considered redshifts. The overall shape of the spectrum remains consistent between the different models and redshift, varying mostly in amplitude. Figure \ref{ps_rat} replots the ratio of each MG power spectrum to the corresponding GR spectrum, for each different redshift.

Figure \ref{ps_rat} shows that the power spectrum of each model is broadly consistent with an overall change in amplitude compared to GR, but little change in shape. The stronger $f(R)$ models display an overall reduction in power spectrum amplitude at all redshifts. The nDGP models, in contrast, show a suppressed power spectrum at high redshift but return close to  GR at $z=0$. Additionally, the nDGP models show a larger power-spectrum scatter at high $k$ than the $f(R)$ models. However, while there was little change in shape were we to calculate the fractional difference, $1 - P_{MG}(k)/P_{GR}(k)$, with catalogues containing more galaxies in the chosen mass range, which would thus give less noisy curves, some change in the shape would be found.

Our results, with a cluster mass cutoff, are similar to those in Figure 4 of \cite{Alam21} (which uses a smaller mass cutoff). Additionally, because large halos tend to cluster more, the amplitude is greater than that of the full matter power spectrum. Even though MG generically boosts the strength of gravity, we see a reduction of power in the case of the $f(R)$ gravity models. While this result is unintuitive, it is consistent with the explanation in Ref.~\cite{Hernandez18}: (1) Since the enhanced strength of gravity is environmentally sensitive, whenever the chameleon screening is efficient we see more halos form in low-density regions compared to GR. This reduced the difference in the density of large halos between high and low-density regions, leading to less power on the scales of large overdensities and voids. (2) Stronger gravity overall leads initially to more halos at a given mass, but that results in a higher merger rate. The combination of these also will modify the power spectrum, but the direction depends on the detailed balance between halo formation and merger rate. However, both of these effects are due to the imposition of a halo mass cutoff. If we were to consider the full matter distribution, we would recover the enhanced matter power spectrum expected if the strength of gravity has been boosted at all scales (cf.\ Fig.~1 of Ref.~\cite{Alam21}).

This leads us to the conclusion that the relationship between the cluster halo power spectrum and the underlying gravity model is complex, and care must be taken in the interpretation of measurements. 

\section{Mean Pairwise Velocity}

\begin{figure}
	\centering
	\includegraphics[width=8.6cm]{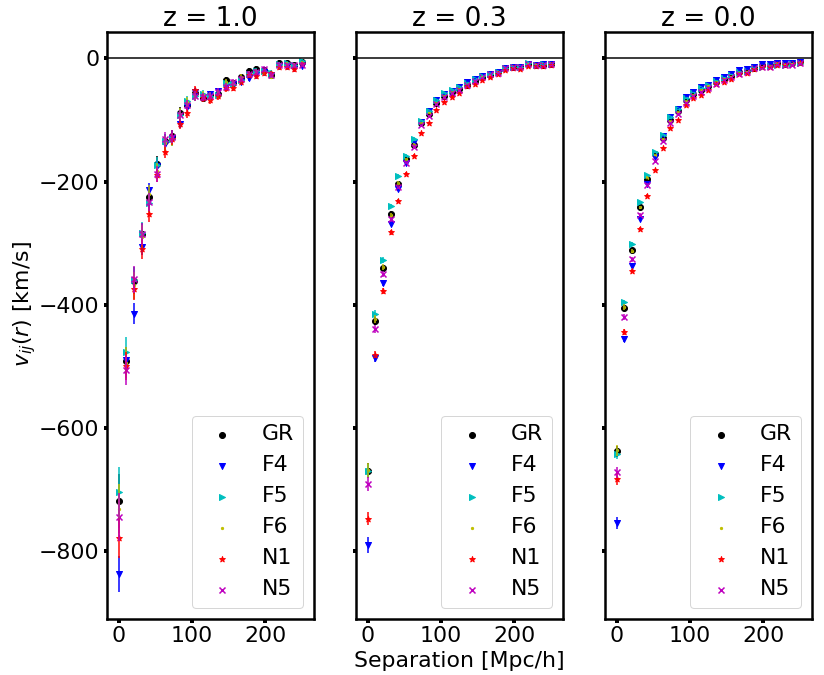}
	\caption{Comparison of the mean pairwise velocity as a function of co-moving separation, in separation bins of 10 Mpc/h, for all gravity models, for redshifts $z = 1$ (left), $z = 0.3$ (middle), $z = 0$ (right). The standard error on the mean is displayed on each point.}
	\label{mpv_all}
\end{figure}

\begin{figure}
	\centering
	\includegraphics[width=8.6cm]{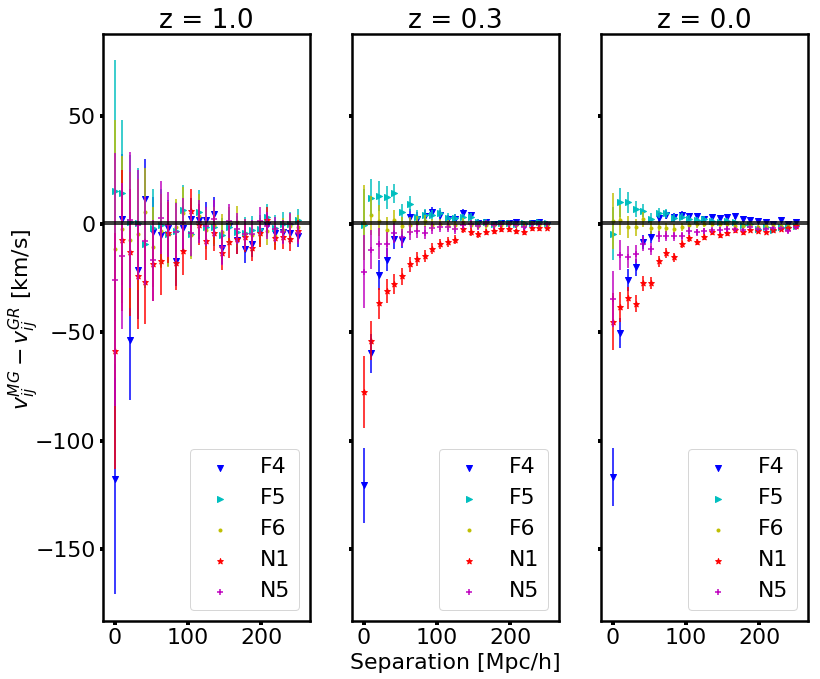}
	\caption{The difference between the mean pairwise velocity for a modified gravity model and GR, in separation bins of 10 Mpc, for redshifts $z = 1$ (left), $z = 0.3$ (middle), $z = 0$ (right).The standard error for each point combines the GR and MG errors in quadrature.}
	\label{mpv_diff}
\end{figure}


We also consider the cluster mean pairwise peculiar velocity, as future microwave background maps may provide a path to a clean measurement of this quantity. The mean pairwise peculiar velocity (MPV) $v(r)$ \cite{Davis77} is the mean of the relative velocity component along the line of separation between each possible pair of cluster halos,
\begin{equation}
    v(r) = \langle\vu{r}_{ij}\cdot (\bv_i - \bv_j)\rangle,
\end{equation}
where $r = \abs{\br_i - \br_j}$ is the magnitude of the co-moving separation between the i-th and j-th cluster, $\vu{r}_{ij}$ given by
\begin{equation}
    \vu{r}_{ij} = \frac{\br_i - \br_j}{|\br_i - \br_j|},
\end{equation}
is the unit vector parallel to the line that connects the i-th and j-th halo, and $\bv_i$ is the velocity of the i-th halo and $\langle\cdots\rangle$ is mean over all possible pairs $i,j$ of clusters. If the distribution of cluster velocities were random, we would anticipate the mean to be zero (clusters would be just as likely to fall to move toward each other as they are to move outward). However, since gravity is an attractive force, we expect to find, on average, a small negative mean velocity for all pair of clusters at a given separation. This makes MPV a promising tool to test GR on scales of tens of Mpc because it directly probes the effect of of gravity on these scales. 

Figure \ref{mpv_all} shows the MPV at each redshift as a function of the co-moving separation between two halos. Normal errors on the sums over pairs are shown. The velocities are all negative, as expected. The models which boost the strength of gravity at small separations tend to show more negative velocities. This can more clearly be seen in Fig.~\ref{mpv_diff} which shows the difference between the MPV in each gravity model and GR. The various models show a range of behaviors. The two nDGP models show consistently higher MPV than general relativity at all separations, while the $f(R)$ models have a less consistent behavior with separation. This is likely due to screening in $f(R)$ depending on the local density, inducing more complex velocity variations. 

Another notable feature is that the MPV decreases with decreasing redshift, due to the effect of late-time acceleration on large-scale structure formation. The differences between models also becomes less pronounced at late times, due to the damping of the velocities by the cosmic acceleration common to all the models. Compared to the power spectrum, the regularity of the mean pairwise velocity and its distinctive evolution with redshift make it a promising diagnostic for any departures from general relativity. 

\section{Summary, Conclusion, and Outlook}

We have analyzed the output of several modified gravity simulations of large-scale structure to extract statistical properties of galaxy clusters. Starting from identical initial conditions, the simulations host corresponding galaxy clusters to the standard gravity case, with some scatter in cluster mass and displacements in cluster positions and velocities which we quantify. We then extract the cluster mass function, power spectrum, and mean pairwise velocity for each simulation. For the simulations we consider, the differences in these statistics range from a few percent to tens of percent. If gravity is modified, differences in these cluster statistics will be correlated, providing leverage in distinguishing modified gravity from systematic errors. 

The models considered, of the DGP or $f(R)$ form, are a range of representative modified gravity models. While a cosmological gravity theory may be different in detail, we expect cluster statistics to be roughly within the range of models considered here. Greater effects would likely have already been spotted, while it is plausible that a modification of gravity substantial enough to generate accelerating expansion will have some non-negligible effect on galaxy cluster scales. Any cosmological modification of gravity must be shielded from significant effects at solar system scales where departures from general relativity are strongly constrained observationally (e.g., \cite{Vikram13}).

Modified gravity constraints on scales of galaxy clusters or larger scales can provide a powerful complement to the usual post-Newtonian constraints on solar system scales. Here we have evaluated three galaxy cluster statistics -- mass function, spatial power spectrum, and mean pairwise velocity --  which show a range of variations from a few percent to tens of percent for the models considered. If a modification of gravity is correct, these three signals are correlated, which gives further leverage in detecting small departures from general relativity. 

Galaxy clusters are particularly interesting candidates for probing subtleties of structure growth, for two reasons: current and upcoming experiments compile large cluster catalogs with well-understood completeness, and cluster catalogs compiled through their thermal Sunyaev-Zeldovich distortions of the cosmic microwave background detect all clusters along the line of sight regardless of redshift. Arcminute-scale temperature decrements at microwave frequencies below the 220 GHz null are an unambiguous cluster signature. The latest arcminute-resolution microwave maps over large sky areas from the Atacama Cosmology Telescope yield a catalog of over 4000 galaxy clusters with masses above $5\times 10^{14} M_\odot$ \cite{Hilton21}, and the next data release will likely double this number to give the largest cluster catalog to date. Cluster SZ signals detected at greater than $5\sigma$ statistical significance have been demonstrated to give essentially a complete mass-limited cluster catalog over a given sky region \cite{Hasselfield13} for clusters at $z > 0.2$; the cluster mass limit therefore decreases as map sensitivity improves. Our cluster sample is anticipated to increase to at least 16,000 clusters with masses down to $2\times 10^{14} M_\odot$ detected at the same significance with the upcoming Simons Observatory \cite{SO19}, and likely many more down to lower cluster mass limits with CMB-S4. 

Masses for clusters can be obtained through the correlation of cluster mass with the thermal SZ signal, although galactic emission from within clusters complicates this route \cite{Dicker21}. Gravitational lensing masses will ultimately be available, either through weak lensing in large optical surveys (see, e.g., \cite{Murata19}) or lensing of the microwave background \cite{Madhavacheril15, Madhavacheril20}. Either of these techniques can be used to compile the cluster mass function. The cluster power spectrum and the mean pairwise velocity as a function of separation require positions in three-dimensional space, which can be obtained from cluster redshifts, either photometric or spectroscopic. The Dark Energy Survey and the upcoming Rubin Telescope will provide photometric redshifts for sky regions largely overlapping with ACT, Simons Observatory, and CMB-S4, while current large spectroscopic survey instruments have demonstrated the capability to obtain great numbers of spectroscopic redshifts. 

Cluster velocities along the line of sight induce the kinematic Sunyaev-Zeldovich temperature shift in the microwave background; this signal has been detected directly via the mean pairwise momentum statistic \cite{Calafut21, Bernardis17, Hand12, Planck16, Soergel16} and by cross-correlating with a cosmic velocity field inferred from the large-scale galaxy distribution \cite{Schaan16}. Extracting cluster velocities from the kSZ signal is complicated by the signal being proportional to the bulk gas momentum rather than velocity; any claimed signature of modified gravity might also arise as a systematic error in cluster gas modeling.  

A cleaner alternative is the transverse velocity lensing signal \cite{Birkinshaw83, Gurvits86}, a dipole CMB temperature distortion which is proportional to the cluster transverse velocity and the total cluster mass. Combined with cluster mass estimates from lensing, this signal holds the promise of a clean velocity probe. Its amplitude is small, and so far undetected, although a first detection is likely with coming experiments \cite{Hotinli19, Yasini19}. 

Given these basic observables, estimators for the galaxy cluster statistics considered in this paper can be formulated, accounting for various noise sources which must be mitigated. SZ-selected galaxy clusters avoid some of the more difficult selection effects associated with many astronomical catalogs. While precise masses of individual clusters from microwave background lensing are far off, statistical mass determinations will rapidly improve, and optical weak lensing masses from LSST (whose sky coverage is an excellent match for ACT, Simons Observatory, and CMB-S4) will leverage significantly greater depth and image quality compared to current optical surveys. Transverse lensing of the microwave background to determine cluster velocities likely has significantly lower astrophysical systematics compared to other velocity probes such as redshift-space distortions or kinematic SZ. The ultimate limiting systematic effects in the galaxy cluster statistics considered here are not well understood; detailed studies will be required to determine the eventual ability of galaxy cluster statistics to constrain modifications of gravity. 

The analysis presented here demonstrates that if one galaxy cluster statistic departs from the predictions of standard cosmological models due to a modification of gravity, other statistics will also display potentially measurable differences. This provides a way to distinguish subtle differences in structure growth due to modified gravity from possible systematic errors in these challenging measurements. Correlations between the statistics computed here in generic modifications of gravity will be considered elsewhere. Other constraints on modified gravity from galaxy clustering and cluster abundances can be found in Ref.~\cite{Liu21}.

An alternate approach to probing structure formation is to construct estimators of the total baryon velocity or density by cross-correlating Fourier modes of the the entire distributions of galaxies and microwave background signals (for example, Ref.~\cite{Hotinli_2021}) rather than using only the galaxy clusters as tracers of the mass distribution. These types of estimators should be equivalent to using all of the halos down to some mass limit below that of galaxy clusters. This technique has the advantage of simplicity and of using all available data; but it also picks out spurious signals from other unrelated signals in the data which must be subtracted through modeling. It remains to be seen whether the increase in data is an advantage over using only galaxy clusters given the increased difficulty in accounting for systematic errors. Ultimately, estimating structure growth using multiple different techniques with different systematics is the best route to robust conclusions about subtle signals such as those from modified gravity. 

The results of our halo-matching exercise shows that a peak-patch description of halos in modified gravity should be equally good as for general relativity. We have shown that large galaxy clusters are in one-to-one correspondence in the modified gravity and general relativity simulations with identical initial conditions. Individual cluster masses differ due to systematic differences in linear growth and random differences in merger histories. This suggests the possibility of constructing rapid approximate calculations of galaxy cluster locations, masses, and velocities as has been done for standard general relativity with the Websky code \cite{Stein20}. This peak-patch code needs to be modified for a different linear growth function, and masses for the resulting cluster-sized halos determined by abundance matching with the kind of modified gravity simulations used in this paper instead of conventional gravity simulations. The abundance matching procedure naturally incorporates astrophysical differences in clusters between general relativity and modified gravity such as tidal effects or different frequency of minor mergers. The linear growth factor can generally be determined (analytically or numerically) for a given gravity theory. So in principle it is straightforward to create Websky-style sky simulations for other gravity theories which are close to general relativity. We will address this possibility in more detail elsewhere.

One other interesting extension of this work would be to use a general parameterization of modified gravity, such as the Parameterized Post-Friedman formulation of modified gravity \cite{Baker13}. Combining a general framework such as this, plus Websky-style simulations of large-scale structure and corresponding microwave background signatures, will provide an efficient method of constructing useful mocks for confronting upcoming cosmological data with modified gravity models. Given the pressing nature of the fundamental physics questions surrounding the accelerating expansion of the universe, we look forward to the near future when new probes of gravity on cosmological scales become available. 

\section*{Acknowledgements}

We thank Baojiu Li and Wojciech Hellwing on behalf of \cite{Cautun18} and \cite{Hellwing17} for making available the \emph{ELEPHANT} and nDGP simulations, respectively.

\bibliography{first_project}

\end{document}